
\magnification = 1200
\hsize = 15 truecm
\vsize = 23 truecm
\baselineskip 20 truept
\voffset = -0.5 truecm
\parindent = 1 cm
\overfullrule = 0pt
\count0 = 1
\footline={\hfil}

\null

\settabs 3 \columns
\+&&  Preprint-KUL-TF-93/16 \cr
\+&&  (Revised version)  \cr
\+&&  hep-th/9310012  \cr
\+&&  September 1993 \cr

\vskip 1 truecm

\centerline
{\bf The Batalin-Vilkovisky formalism with the Virasoro symmetry}

\vskip 3 truecm

\centerline{\bf S. Aoyama}
\smallskip

\vskip 0.5 cm

\centerline{\bf Instituut voor Theoretische Fysica}
\smallskip
\centerline{\bf Katholieke Universiteit Leuven}
\smallskip
\centerline{\bf Celestijnenlaan 200D}
\smallskip
\centerline{\bf B-3001 Leuven, Belgium}

\vskip 2.5 truecm

\noindent
{\bf Abstract}

We introduce the Virasoro symmetry in the BV formalism and give an explicit
construction of the anti-bracket, which is Virasoro invariant. It is shown that
the master equation with this anti-bracket has an infinite number of solutions.
The base space of the BV formalism is a fermionic version of the Virasoro
manifold $Diff(S^1)/S^1$. We discuss also the Ricci tensor of this fermionic
manifold.

\vskip 2 cm
\noindent
e-mail: shogo\%tf\%fys@cc3.kuleuven.ac.be

\vfill\eject

\footline={\hss\tenrm\folio\hss}

\noindent
1.~~~~~~There is much interest in the BV formalism  going beyond the original
purpose to BRST quantize the gauge theory$^{[1]}$. Its extended viability has
been proved by the recent applications to the non-critical string$^{[2]}$ and
the string field theory$^{[3]}$. The geometry of the formalism has been
considerably clarified in refs 4 and 5.

Recently it has been discussed $^{[6,7]}$ that
the fermionic symplectic structure of the BV formalism can be given by a
fermionic K\"ahler 2-form as a special case.
The base space of the BV formalism  with this symplectic structure is  the so
called fermionic K\"ahler manifold, which is a fermionic version of the usual
K\"ahler manifold.
The fermionic K\"ahler 2-form has been determined by introducing an
isometry$^{[6]}$.
The anti-bracket defined with such 2-form is invariant by an isometry
transformation. Some interesting solutions of the master equation have been
discussed in this case.

The Virasoro  symmetry is an interesting isometry to study in this regard. In
this note we show that the BV formalism can incorporate the Virasoro symmetry
as well, and construct the anti-bracket which is invariant by the Virasoro
transformation.
Since  we deal with an infinite dimensional algebra, the machinery developed in
ref. 6 is not applicable.
Some years ago non-linear realization of the Virasoro $^{[8]}$ algebra was
studied by using the CCWZ  formalism$^{[9]}$. They have shown that the quotient
of the Virasoro group by its one parameter central group is indeed the K\"ahler
manifold, called the Virasoro manifold.
(It is commonly denoted by $Diff(S^1)/S^1$.) The idea is that we use the
technique in ref. 8 to study the corresponding fermionic K\"ahler manifold. An
explicit construction of this fermionic Virasoro manifold is given. It may be
used as a base space of the BV formalism. Then the anti-bracket  is Virasoro
invariant.
We study the master equation of the BV formalism in this case
and discover an infinite number of Virasoro invariant solutions.

In ref. 8 they calculated the Ricci tensor of the bosonic Virasoro manifold:
$$
R_{\alpha \underline \beta} = -{26 \over 12}(\alpha^3 - {1\over 13}\alpha)
\delta_{\alpha + \beta, 0}               \eqno (1)
$$
with $\alpha, \beta = \pm 1, \pm 2, \cdots$. (The meaning of the indices will
be clear in the text.) The curious coincidence between this Ricci tensor and
the Virasoro anomaly raised vivid interest at that time.
It was originally discovered by Bowick and Rajeev$^{[10]}$. Namely they
calculated a curvature of the holomorphic vector bundle over the (bosonic)
K\"ahler manifold $Diff(S^1)/S^1$, in which the fibre is either a string Fock
space or simply the vacuum
 of the Fock space. In this note we examine the  Ricci tensor (1) for the
fermionic Virasoro manifold. We find that it is vanishing.

\vskip 1 cm

\noindent
2.~~~~~~To start with, we shall recall the basic formulae of the symplectic
geo\-metry$^{[3,5]}$. Consider a $2D$ manifold parametrized by coordinates $y^i
= (\phi^1, \phi^2,$  $ \cdots,\phi^D, \xi^1, \xi^2, \cdots,  \xi^D)$ with
$\phi$'s and $\xi$'s bosonic and fermionic respectively. Suppose that it has a
symplectic structure given by a non-degenerate 2-form
$$
\omega = dy^j \wedge dy^i \omega_{ij},         \eqno (2)
$$
which is fermionic and closed
$$
d \omega = 0.
$$
These equations read in components
$$
(-)^{ik}\partial_i \omega_{jk} + (-)^{ji}\partial_j \omega_{ki} +
(-)^{kj}\partial_k \omega_{ij} = 0,        \eqno (3)
$$
$$
\omega_{ij} = -(-)^{ij}\omega_{ji}.        \eqno (4)
$$
Here we have used the short-hand notation for the grassmannian parity of the
coordinates $\varepsilon (y^i) = i$ in the sign factor. By this notation  we
have
$\varepsilon (\omega_{ij}) = i + j + 1$. Define the inverse of $\omega_{ij}$ by
$$
\omega_{ij}\omega^{jk} = \omega^{kj}\omega_{ji} = \delta_i^k. \eqno (5)
$$
Then eqs (3) and (4) may be written respectively as
$$
(-)^{(i+1)(k+1)}\omega^{il}\partial_l \omega^{jk} +
(-)^{(j+1)(i+1)}\omega^{jl}\partial_l \omega^{ki} +
(-)^{(k+1)(j+1)}\omega^{kl}\partial_l \omega^{ij} = 0,
$$
$$
\omega^{ij} = -(-)^{(i+1)(j+1)}\omega^{ji}.
$$
With this fermionic symplectic structure the anti-bracket of
the BV formalism is given by
$$
\{A,B\} = (-)^{i[\varepsilon (A) + 1]}\partial_i A \omega^{ij} \partial_j B,
\eqno (6)$$
We also define a nilpotent second order differential operator by
$$
\Delta \equiv {1 \over \rho }(-)^i \partial_i(\rho \omega^{ij}\partial_j).
\eqno (7)
$$
They are  related with each other by
$$
\Delta (AB) = \Delta A \cdot B + (-)^{\varepsilon (A)} A\Delta B  +
(-)^{\varepsilon (A) } \{A,B \}.  \eqno (8)
$$
The operator (7) is nilpotent if
$$
\Delta[{1 \over \rho} (-)^i \partial_i(\rho \omega^{ij} )] = 0.  \eqno (9)
$$

We may introduce an isometry in the manifold. It is realized by a set of
Killing vectors $V^{Ai}(y)$, $A = 1,2,\cdots,N$,  which obey the Lie algebra of
a group $G$
$$
V^{Ai}\partial_i V^{Bj} - V^{Bi}\partial_i V^{Aj} = f^{ABC}V^{Cj},
$$
with structure constants $f^{ABC}$. The grassmannian parities  are assigned as
$\varepsilon (V^{Ai})$ $ = i$.
Then the fermionic symplectic structure $\omega_{ij}$ satisfies the Killing
condition
$$
{\cal L}_{V^A}\omega_{ij}  \equiv V^{Ak}\partial_k \omega_{ij} + \partial_i
V^{Ak}\omega_{kj}- (-)^{ij}\partial_j V^{Ak}\omega_{ki} = 0.  \eqno (10)
$$
In terms of the inverse $\omega^{ij}$ this condition becomes
$$
{\cal L}_{V^A}\omega^{ij}  \equiv V^{Ak}\partial_k \omega^{ij} -
\omega^{ik}\partial_k V^{Aj} + (-)^{(i+1)(j+1)}\omega^{jk}\partial_k V^{Ai}
= 0.   \eqno (11)
$$
We may find an explicit form of $\omega_{ij}$  as a simultaneous solution of
eqs (3) and (10). Owing to the Killing condition (11),
the anti-bracket (6) is invariant by the isometry transformations given by the
Killing vectors
$$
\delta y^i = \epsilon^A V^{Ai},
$$
in which $\epsilon^A$ are global parameters. In ref. 6 this program has been
worked out by extending the isometry of the hermitian symmetric space

\vskip 1 cm

\noindent
3.~~~~~~Assuming that it  is infinite dimensional ($D = \infty$), we can
introduce also the Virasoro symmetry in the manifold. The resulting manifold is
a fermionic version of the Virasoro manifold $Diff (S^1)/S^1$ discussed in ref.
8. It is convenient
 to summarize the CCWZ formalism for the bosonic Virasoro manifold, since we
will construct the fermionic one  on that basis.

Consider the Virasoro algebra
$$
[L_a , L_b ] = (a - b)L_{a + b },     \eqno (12)
$$
without the anomaly. The Virasoro manifold $Diff (S^1)/S^1$ is the quotient of
the Virasoro group by its one parameter central group generated by $L_0$. A
standard way to parametrize this coset space is to write a Virasoro group
element
$$
g = \exp (i\sum_{ a, \alpha \ne 0} \phi^\alpha L_a \delta^a_\alpha),   \eqno
(13)
$$
where $\phi^\mu$ can be used as coordinates of the manifold $Diff(S^1)/S^1$.
The generators $L_a$ satisfy the hermitian condition $L_a^{\ \dag} = L_{-a}$,
so that the manifold admits a complex structure as
$$
(\phi^\alpha )^* = \phi^{-\alpha }.      \eqno (14)
$$
The Cartan-Maurer 1-form is defined by
$$
g^{-1}d g = \sum_{a} e^a L_a.
$$
By exterior differentiation we get the Cartan-Maurer equation
$$
de^a = -{1 \over 2}\sum_{ b,c}(b-c)\delta_{b+c}^a \ e^b e^c.     \eqno (15)
$$
In components it reads
$$
{ \partial \over \partial \phi^\alpha} e^a_\beta
- { \partial \over \partial \phi^\beta} e^a_\alpha
 = -\sum_{ b,c}(b-c)\delta_{b+c}^a \ e^b_\alpha e^c_\beta.   \eqno (16)
$$
When multiplied from the left by an element of the Virasoro group, the group
element (13) transforms as
$$
e^{i\sum_{a} \epsilon^a L_a}\cdot  g = \exp (i \sum_{a,\alpha \ne 0}
\Phi^\alpha (\phi)L_a \delta^a_\alpha)\cdot h.     \eqno (17)
$$
Here $\epsilon^a$ are global parameters of the transformation and $h$ is the
so-called compensator
$$
h = e^{i\lambda (\phi) L_0},
$$
with an appropriate function $\lambda (\phi)$. This defines non-linear
transformation of the coordinates
$$
\phi^\alpha \longrightarrow  \Phi^\alpha (\phi) = \phi^\alpha + \sum_a
\epsilon^a R_a^{\ \alpha} (\phi) + O((\epsilon^a)^2),    \eqno (18)
$$
in which $R_a$ are the Killing vectors of the manifold. Under this  the
coefficients $e^a, \  (a \ne 0)$, transform as
$$
e^a \longrightarrow e^{-i\lambda (\phi) a } e^a = (1 - i\lambda (\phi) a +
O(\lambda^2))e^a,         \eqno (19)
$$
(no sum over $a$).

The Virasoro manifold $Diff(S^1)/S^1$ has the (bosonic) symplectic structure
$$
\Omega_{\alpha \beta} = \sum_a f(a)e^a_\alpha e^{-a}_\beta, \eqno (20)
$$
in which
$$
f(a) = A a^3 + B a,   \eqno (21)
$$
with arbitrary constants $A$ and $B$. Indeed we can check that it satisfies
$$
{\partial \over \partial \phi^\alpha} \Omega_{\beta \gamma} + {\partial \over
\partial \phi^\beta} \Omega_{\gamma \alpha}
+ {\partial \over \partial \phi^\gamma} \Omega_{\alpha \beta} = 0,  \eqno (22)
$$
by using the Cartan-Maurer equation (16). Having the complex structure given by
eq. (14) the Virasoro manifold $Diff(S^1)/S^1$ is a K\"ahler manifold.

\vskip 1 cm

\noindent
4.~~~~~~So far we have summarized the CCWZ formalism for the Virasoro manifold
$Diff(S^1)/S^1$\ $^{[8]}$. We now consider a new manifold by introducing
fermionic coordinates $\xi^\alpha $ corresponding to $\phi^\alpha $, with
$$
(\xi^\alpha)^* = \xi^{-\alpha}.   \eqno (23)
$$
We associate the transformation law
$$
\xi^\alpha \longrightarrow \xi^\beta {\partial \over \partial \phi^\beta }
\Phi^\alpha (\phi).
$$
The Killing vectors of the new manifold are given by
$$
V_a^{\ i} = (R_a^{\ \alpha}, \xi^\beta {\partial \over \partial \phi^\beta }
R_a^{\ \alpha}).  \eqno (24)
$$
Consider the following matrix
$$
\eqalignno{
\omega_{ij} & = \left(
\matrix{ \omega_{\phi \phi} & \omega_{\phi \xi}  \cr
         \omega_{\xi \phi} & \omega_{\xi \xi}  \cr } \right)  &  \cr
&       &     \cr
& = \left(
\matrix{ \xi^\gamma {\partial \over \partial \phi^\gamma }
 \Omega_{\alpha \beta}
 & \Omega_{\alpha \beta}   \cr
 &      \cr
 \Omega_{\alpha \beta}   &   0   \cr}\right),  &  (25) \cr}
$$
with eq. (20). First of all the symmetric property (4) is evident. Secondly
this $\omega_{ij}$ satisfies eq. (3)  by means of eq. (22).
Finally it satisfies also the Killing condition (10). For instance the Killing
condition for the block matrix
$\omega_{\phi \phi}$ reads
$$
\eqalign{
& V_b^{\ i}\partial_i  \{ \sum_a f(a)e^a_{[\alpha} \xi^\gamma {\partial \over
\partial \phi^\gamma } e^{-a}_{\beta ]}\}     \cr
&  = -[ {\partial \over \partial \phi^\alpha}R_b^{\ \delta} \sum_a
f(a)e^a_{[\delta} \xi^\gamma {\partial \over \partial \phi^\gamma }
e^{-a}_{\beta ]}
 + {\partial \over \partial \phi^\alpha}{\partial \over \partial
\phi^\gamma}R_b^{\ \delta} \cdot \xi^\gamma \sum_a f(a)e^a_\delta e^{-a}_\beta
]  \cr
&  \quad \quad \quad \quad \quad
 + [ \alpha \rightleftharpoons
\beta ].   \cr}  \eqno (26)
$$
It can be shown as follows. Note that
$$
R_b^{\ \delta}{\partial \over \partial \phi^\delta}e^a_\alpha = -{\partial
\over \partial \phi^\alpha}R_b^{\ \delta}\cdot e^a_\delta - i\lambda a\
e^a_\alpha,
$$
which follows from eq. (19) together with eq. (18). By using this the l.h.s. of
eq. (26) is calculated as
$$
\eqalign{
-\sum_a f(a)[ & {\partial \over \partial \phi^\alpha}R_b^{\ \delta}\cdot e^a_
\delta
\xi^\gamma {\partial \over \partial \phi^\gamma}e^{-a}_\beta
 + {\partial \over \partial \phi^\beta }R_b^{\ \delta}\cdot e^a_\alpha
\xi^\gamma {\partial \over \partial \phi^\gamma}e^{-a}_\delta  \cr
& + \xi^\gamma {\partial \over \partial \phi^\gamma}{\partial \over \partial
\phi^\beta}R_b^{\ \delta}\cdot e^a_\alpha e^{-a}_\delta] + ( \alpha
\rightleftharpoons \beta).
\cr }
$$
Here we would like to remark that the $\lambda$-dependent pieces disappeared.
This is exactly the r.h.s. of eq. (26). The other part of the Killing condition
can be easily checked. Thus we have obtained the closed fermionic 2-form (2)
with eqs (25). The manifold with the symplectic structure given by this 2-form
is a fermionic version of the Virasoro manifold $Diff(S^1)/S^1$ discussed in
ref. 8, called the fermionic Virasoro manifold.
It is a K\"ahler manifold , since we have the complex structure by eqs (14) and
(23). The matrix (25) can be  inverted as
$$
\eqalignno{
\omega^{ij} & = \left(
\matrix{ \omega^{\phi \phi} & \omega^{\phi \xi}  \cr
         \omega^{\xi \phi} & \omega^{\xi \xi}  \cr } \right)  &  \cr
&       &     \cr
& = \left(
\matrix{  0  & \Omega^{\alpha \beta}   \cr
 &      \cr
 \Omega^{\alpha \beta}   &
 \xi^\gamma {\partial \over \partial \phi^\gamma } \Omega^{\alpha \beta }
\cr}\right),  &  (27) \cr}
$$
by eq. (5). Here $\Omega^{\alpha \beta}$ is the inverse of the block matrix
$\Omega_{\alpha \beta}$:
$$
\Omega_{\alpha \beta}\Omega^{\beta \gamma} = \Omega^{\gamma \beta}\Omega_{\beta
\alpha} = \delta^\gamma_\alpha.
$$
(An explicit form of $\Omega^{\alpha \beta}$ can be found by solving  these
equations perturbatively arround the origin of the coordinates.)
Owing to the Killing condition (11) the anti-bracket (6)  with eq. (27) is
invariant by the Virasoro transformations given by the Killing vectors (24):
$$
\delta y^i = \sum_a \epsilon^a V_a^{\ i}. \eqno (28)
$$

\vskip 1 cm

\noindent
5.~~~~~~ We may be interested in solving the master equation of
the BV formalism
$$
\Delta \Psi = 0,
$$
with $\Delta$ defined by eq. (7) where the symplectic structure $\omega^{ij}$
is given by eq. (27). It can be written as
$$
\Delta_0 \Psi \ + \ \{\log \rho, \Psi \} = 0,
$$
in which $\Delta_0$ is the normal-ordered form of $\Delta$
$$
\Delta_0 = (-)^i \omega^{ij} \partial_j \partial_i, \eqno (29)
$$
and  the  second piece is the anti-bracket defined by eq. (6).
By eq. (8) we further calculate the l.h.s. to find
$$
{1 \over \rho} \Delta_0 (\rho \Psi) - {1 \over \rho} (\Delta_0 \rho ) \Psi = 0.
 \eqno (30)
$$
The normal-ordered operator $\Delta_0$ with eq. (27) is nilpotent by using
$$
\Omega^{\alpha \eta}{ \partial \over \partial \phi^\eta } \Omega^{\beta \gamma}
+
\Omega^{\beta \eta}{ \partial \over \partial \phi^\eta } \Omega^{\gamma
\alpha}+
\Omega^{\gamma \eta}{ \partial \over \partial \phi^\eta } \Omega^{\alpha \beta}
= 0,
$$
which follows from eq. (22). If we require that
$$
\Delta_0 \rho = 0,  \eqno (31)
$$
$\Delta$ is also nilpotent$^{[5]}$. This is a stronger condition than eq. (9).
Then eq. (30) becomes
$$
\Delta_0 \Psi_\rho = 0,  \eqno (32)
$$
with $\Psi_\rho = \rho \Psi$.
We find that $\Psi_\rho$ and $\rho$ obey the same master equation with
$\Delta_0$. A special solution to eq. (31) or (32) is given by
$$
S_0 = \sum_a \Omega_{\alpha \beta}\xi^\alpha \xi^\beta =  \sum_a f(a)
e^a_\alpha e^{-a}_\beta \xi^\alpha \xi^\beta,
$$
which is invariant by the Virasoro transformations (28).
This solution generates  infinitely many other solutions  such that
$$
S = \sum_{\alpha = 0}^\infty c_\alpha (S_0)^\alpha, \eqno (33)
$$
with arbitrary constants $c_\alpha$.

\vskip 1 cm

\noindent
6.~~~~~~The rest of this note is dedicated to calculation of the Ricci tensor
of the fermionic Virasoro manifold. It can be done quite similarly to the
bosonic  case$^{[8]}$. First of all we have to make the technique elaborated in
ref. 8 applicable for calculating the Ricci tensor of the fermionic K\"ahler
manifold. The fermionic K\"ahler manifold is a complex supermanifold which can
be
 parametrized by  holomorphic supercoordinates
$$
{\bf z}^\mu = (z^\alpha,\ \zeta^\alpha), \quad \quad \alpha = 1,2, \cdots, D,
$$
and their complex conjugates with $z$'s and $\zeta$'s bosonic and fermionic
respectively. It has a symplectic structure given by the closed 2-form
$$
\omega = i d \overline {\bf z}^{\underline \nu }\wedge  d  {\bf z}^{ \mu }
\gamma_{\mu \underline  \nu}.
$$
Here $\gamma_{\mu \underline \nu}$ is the fermionic metric of the manifold,
i.e., the grassmannian parity is $\varepsilon (\gamma_{\mu \underline \nu}) =
\varepsilon (\mu) + \varepsilon (\nu) + 1$. This 2-form is a special case of
the general one given by eq. (2). There exist a fermionic K\"ahler potential
such that
$$
\gamma_{\mu \underline \nu} = \partial_\mu \partial_{\underline \nu} K.
$$
The inverse metric $\gamma^{\mu \underline \nu}$ is defined by
$$
\gamma_{\mu \underline \nu}\gamma^{\underline \nu \eta} = \gamma^{\eta
\underline \nu}\gamma_{\underline \nu \mu} = \delta_\mu^\eta,  \eqno (34)
$$
and satisfies
$$
\gamma^{\mu \underline \nu} = (-)^{(\mu + 1)(\nu + 1)}\gamma^{\underline \nu
\mu}.
$$
The affine connections are given by$^{[6]}$
$$
\Gamma^{\ \ \ \eta}_{\mu \nu}  = \partial_\mu \gamma_{\nu \underline \rho}\cdot
\gamma^{\underline \rho \eta},  \quad \quad
\Gamma^{\ \ \ \underline \eta}_{\underline \mu\underline
\nu}  = \partial_{\underline \mu} \gamma_{\underline \nu  \rho}\cdot
\gamma^{\rho \underline \eta },
$$
and  other components are vanishing. The covariant derivative for vectors is
defined by
$$
\eqalign{
D_\mu A_\nu  = \partial_\mu A_\nu - \Gamma^{\ \ \ \rho}_{\mu \nu}A_\rho, &
\quad \quad
D_\mu A^\nu  = \partial_\mu A^\nu + A^\rho \Gamma^{\ \ \ \nu}_{\rho \mu},  \cr
&  {\rm c.c..} \cr}
$$
The curvature tensor is given by
$$
\eqalign{
R_{\underline \nu \mu \sigma}^{\ \ \ \ \eta}  = \partial_{\underline \nu}
\Gamma^{\ \ \ \eta}_{\mu \sigma} , & \quad \quad
R_{\mu \underline \nu  \sigma}^{\ \ \ \ \eta}  = -(-)^{\mu
\nu}\partial_{\underline
\nu} \Gamma^{\ \ \ \eta}_{\mu \sigma},  \cr
&   {\rm c.c..}   \cr}
     \eqno (35)
$$
Other components are vanishing. We obtain the Ricci tensor as
$$
R_{\mu \underline \nu} = -(-)^{\mu \nu}R_{\underline \nu \mu} =
 -(-)^{\mu \nu + \sigma} R_{\underline \nu \mu \sigma}^{\ \ \ \ \sigma}.
\eqno (36)
$$

The fermionic K\"ahler manifold can admit an isometry. It is realized by a set
of holomorphic Killing vectors ${\bf R}^{A \mu}({\bf z})$ and their complex
conjugates $\overline {\bf R}^{A \underline \mu}(\overline {\bf z})$. They
satisfy the Lie algebra
$$
{\bf R}^{A \mu}\partial_\mu {\bf R}^{B \nu} -
{\bf R}^{B \mu}\partial_\mu {\bf R}^{A \nu} =
f^{ABC}{\bf R}^{C \nu},            \eqno (37)
$$
and the Killing condition
$$
{\cal L}^A \gamma_{\mu \underline \nu} \equiv [{\bf R}^{A \rho}\partial_\rho
 + {\bf R}^{A \underline \rho}\partial_{\underline \rho}]
\gamma_{\mu \underline \nu} + \partial_\mu {\bf R}^{A \rho} \gamma_{\rho
\underline \nu } + (-)^{\mu \nu}\partial_{\underline \nu }\overline {\bf R}^{A
\underline \rho}\gamma_{\underline \rho \mu } = 0,     \eqno (38)
$$
or equivalently
$$
\eqalign{
{\cal L}^A \gamma^{\mu \underline \nu}  \equiv [{\bf R}^{A \rho}\partial_\rho
 + {\bf R}^{A \underline \rho}\partial_{\underline \rho}]
\gamma^{\mu \underline \nu}\ \ -\ \ & \gamma^{\mu \underline
\rho}\partial_{\underline \rho}\overline {\bf R}^{A\underline \nu}    \cr
& -(-)^{(\mu + 1)(\nu + 1)}\gamma^{\underline \nu \rho}\partial_\rho {\bf R}^{A
\mu}
 = 0,    \cr}   \eqno (39)
$$
due to eq. (34).
It is worth checking that the Ricci tensor given by eq. (36) is indeed
covariant by the transformations
$$
\delta {\bf z}^\alpha = \epsilon^A {\bf R}^{A \alpha}({\bf z}), \quad \quad
{\rm c.c.},
$$
with global parameters $\epsilon^A$. Note that the sign factor $(-)^\sigma$  in
eq. (36) does a right work for this. A little calculation shows that eq. (38)
can be written as
$$
\partial_\mu ({\bf R}^{A \rho}\gamma_{\rho \underline \nu }) + (-)^{\mu
\nu}\partial_{\underline \nu}(\overline {\bf R}^{A \underline
\rho}\gamma_{\underline \rho \mu}) = 0 \quad \quad ({\rm Killing\ equation}).
\eqno (40)
$$
By multiplying by the Killing vectors, the curvature tensor takes the form
$$
\eqalignno{
(R^{A B})^{\ \eta}_{\sigma} & \equiv ({\bf R}^{B\mu}\overline {\bf
R}^{A\underline \nu} -
{\bf R}^{A\mu}\overline {\bf R}^{B\underline \nu})R_{\underline \nu \mu
\sigma}^{\ \ \ \ \eta}   &          \cr
& = (D^{[A} D^{B]} - f^{ABC}D^C )^{\ \eta}_{\sigma},   &  (41)  \cr}
$$
in which
$$
D^A \equiv {\bf R}^{A\mu}D_\mu + \overline {\bf R}^{A\underline
\mu}D_{\underline \mu}.
$$
We consider the difference operator
$$
\varphi^A \equiv  {\cal L}^A - D^A.
$$
It is important to note that it does not contain any derivative and  operates
as a matrix on tensors. For instance on a tensor $T_{\mu \underline \nu}$  we
have
$$
\eqalign{
\varphi^A T_{\mu \underline \nu} & \equiv
[({\bf R}^{A \rho}\partial_\rho
 + {\bf R}^{A \underline \rho}\partial_{\underline \rho})
T_{\mu \underline \nu} + \partial_\mu {\bf R}^{A \rho} T_{\rho \underline \nu }
+ (-)^{\mu (\nu +\rho)}\partial_{\underline \nu }\overline {\bf R}^{A
\underline \rho}T_{\mu \underline \rho  }]     \cr
& -[({\bf R}^{A \rho}\partial_\rho
 + {\bf R}^{A \underline \rho}\partial_{\underline \rho})
T_{\mu \underline \nu} - {\bf R}^{A\rho}\Gamma^{\ \ \ \sigma}_{\rho
\mu}T_{\sigma \underline \nu} - (-)^{\mu
(\nu +\sigma)}\overline {\bf R}^{A\underline \rho} \Gamma^{\ \ \ \underline
\sigma}_{\underline \rho \underline \nu}T_{\mu \underline \sigma}]   \cr
& = (\varphi^A)_\mu^{\ \rho}T_{\rho \underline \nu} + (-)^{\mu(\nu + \rho)}
(\varphi^A)_{\underline \nu }^{\ \underline \rho} T_{\mu \underline \rho},
\cr}
$$
with
$$
(\varphi^A)_\mu^{\ \rho} = - D_\mu {\bf R}^{A \rho},  \quad \quad
(\varphi^A)_{\underline \mu}^{\ \underline \rho} = - D_{\underline
\mu}\overline {\bf R}^{A \underline \rho}.   \eqno (42)
$$
Due to the Killing equation (40) the matrices are related by
$$
(\varphi^A)_\mu^{\ \rho}\gamma_{\rho \underline \nu} = -(-)^{\mu \nu}
(\varphi^A)_{\underline \nu}^{\ \underline \rho}\gamma_{\underline \rho \mu},
 \eqno (43)
$$
which will be useful later. The curvature tensor given by (41) can be expressed
 in terms of the difference operator
$$
(R^{AB})^{\ \eta}_{\sigma}  = (\varphi^{[A} \varphi^{B]} - f^{ABC}\varphi^C
)^{\  \eta}_{\sigma},  \eqno (44)
$$
by using the formulae
$$
[{\cal L}^A, {\cal L}^B] = f^{ABC} {\cal L}^C,  \quad \quad  [{\cal L}^A, D^B]
= f^{ABC}D^C.
$$

\vskip 1 cm

\noindent
7.~~~~~~
We now come back to the fermionic Virasoro manifold discussed previously. It
has been shown that it is a K\"ahler manifold. However the coordinates
$(\phi^\alpha, \xi^\alpha)$ and $(\phi^{-\alpha}, \xi^{-\alpha})$, $\alpha >
0$, are mixed under the transformation (17), so that they can not be identified
with the holomorphic coordinates $(z^\alpha, \zeta^\alpha)$ and $(\overline
z^{\underline \alpha}, \overline \zeta^{\underline \alpha})$ discussed just
above. In order to get the holomorphic coordinates we further decompose the
group element (13) as
$$
\eqalign{
g & = \exp (i\sum_{ a, \alpha \ne 0} \phi^\alpha L_a \delta^a_\alpha)  \cr
  & = \exp(i\sum_{a>0} z^\alpha L_a \delta^a_\alpha)
\exp(i\sum_{a>0} w^\alpha L_{-a} \delta^a_\alpha)
\exp(u L_0).   \cr} \eqno (45)
$$
The product of the last two exponentials is an element of the subgroup
generated by $L_a, \  a \le 0$.  By requiring the two expressions of $g$ to
match, $w^\alpha$ and $u$ are found as functions of $z^\alpha$ and $\overline
z^{\underline \alpha}$. They are calculated as power series
$$
\eqalign{
w^\alpha & = \overline z^{\underline \alpha} + \cdots   ,  \cr
u & = \sum_{\alpha >0} \alpha |z^\alpha|^2 + \cdots  .      \cr}
$$
We multiply eq. (45) by a group element as has been done in eq. (17). Then
 $z^\alpha$ transforms holomorphically:
$$
z^\alpha \longrightarrow  \Phi'^\alpha (z) = z^\alpha + \sum_a \epsilon^a
R_a'^{\  \alpha}(z) + O((\epsilon^A)^2).
$$
The transformation law of $\overline z^{\underline \alpha}$ is obtained by
taking complex conjugation of it. (For details we would like to refer to ref.
8.) Correspondingly to $z^\alpha$ and $\overline z^{\underline \alpha}$,
$\alpha > 0$ we introduce  fermionic coordinates $\zeta^\alpha$ and $\overline
\zeta^{\underline \alpha}$ with the transformation law
$$
\zeta^\alpha \longrightarrow \zeta^\beta {\partial \over \partial z^\beta }
\Phi'^\alpha (z).
$$
The supercoordinates ${\bf z}^\mu = (z^\alpha, \zeta^\alpha)$ and $\overline
{\bf z}^{\underline \mu} = (\overline z^{\underline \alpha}, \overline
\zeta^{\underline \alpha})$ can be taken as the holomorphic coordinates  of the
fermionic Virasoro manifold.

In these new coordinates the Killing vectors (24) become holomorphic, i.e.,
$$
{\bf R}_a^{\ \mu} = (R_a'^{\ \alpha}(z), \zeta^\beta {\partial \over \partial
z^\beta } R_a'^{\ \alpha}(z)),   \eqno (46)
$$
and their complex conjugates
$$
({\bf R}_a^{\ \mu} ({\bf z}))^* \equiv \overline {\bf R}_{-a}^{\ \ \ \underline
\mu} (\overline  {\bf z}).    \eqno (47)
$$
The Lie algebra (37) reads
$$
{\bf R}_a^{\ \mu}\partial_\mu {\bf R}_b^{\ \nu}-
{\bf R}_b^{\ \mu}\partial_\mu {\bf R}_a^{\ \nu}
= -i(a-b){\bf R}_{a+b}^{\ \ \ \ \nu}.
$$
The explicit form of (46) may be found in a power series of $z^\alpha$ and
$\overline z^{\underline \alpha}$:
$$
\eqalign{
R_a'^{\ \alpha} & = \delta_a^\alpha +{i \over 2}(2a - \alpha)z^{\alpha - a} +
\cdots , \quad \quad  (a > 0 ),  \cr
R_0'^{\ \alpha} & = - i \alpha z^\alpha + \cdots,   \cr
R_{-a}'^{\ \ \ \alpha} & = - i(2a + \alpha)z^{\alpha + a} + \cdots ,  \quad
\quad (a > 0), \cr}
  \eqno (48)
$$
where no sum is taken over $a$ and
 $z^\alpha = 0$ for $\alpha \le 0$.$^{[8]}$

For the fermionic Virasoro manifold the curvature tensor (44) reads
$$
\eqalign{
(R_{-a b})^{\ \sigma}_\rho  &  = (\varphi_{-a} \varphi_b - \varphi_b
\varphi_{-a}
 - i(a + b) \varphi_{-a + b} )^{\ \sigma}_\rho, \cr
    (a,b,\sigma, \rho & > 0).    \cr}  \eqno (49)
$$
We shall evaluate it at the origin, $z^\alpha = \zeta^\alpha = 0$. This is
sufficient to calculate the Ricci tensor (1) for the fermionic Virasoro
manifold, since it can be determined everywhere by the isometry of the
manifold. (Hereafter all the calculations will be valid in the neighbourhood of
the origin.)
The difference operator $\varphi_{-a}, a > 0$, can be evaluated by
means of (42)  with (48) :
$$
\eqalignno{
(\varphi_{-a})_\mu^{\ \nu} & = \left(
\matrix{ (\varphi_z^{\ z})_\alpha^{\ \beta}  &
         (\varphi_z^{\ \zeta})_\alpha^{\ \beta}   \cr
         &       \cr
         (\varphi_\zeta^{\ z})_\alpha^{\ \beta}  &
         (\varphi_\zeta^{\ \zeta})_\alpha^{\ \beta}  \cr}\right)  &    \cr
&     &    \cr
& = \left(
\matrix{
 i\delta_\alpha^{\ a + \beta}(2a + \beta) &  0  \cr
 &   \cr
 0  &  i\delta_\alpha^{\ a + \beta}(2a + \beta)  \cr}\right),  &  (50)  \cr}
$$
while the difference operator $\varphi_a, a > 0$, can be computed by using eqs
(43), (47) and the above result:
$$
(\varphi_a)_\mu^{\ \nu} = \left(
\matrix{
if(\beta - a)\delta_\alpha^{\ \beta - a}{1 \over f(\beta)}(a + \beta)  &   0
\cr
&   \cr
0  &  -if(\beta - a)\delta_\alpha^{\ \beta - a}{1 \over f(\beta)}(a + \beta)
\cr}\right).  \eqno (51) $$
Here we have known the metric $\gamma_{\mu \underline \nu}$ at the origin as
$$
\gamma_{\mu \underline \nu} = \left(
\matrix{
0  & f(\alpha) \delta_{\alpha \beta}     \cr
&     \cr
f(\alpha)\delta_{\alpha \beta}  &  0   \cr} \right)
$$
from eq. (25) with eqs (20) and (21). These difference operators are identical
with those obtained for the bosonic Virasoro manifold except for the doubling
due to the fermionic coordinates. Therefore
the Ricci tensor corresponding to eq. (1) can be calculated closely following
ref. 8.
Namely with eqs (50) and (51) we calculate the r.h.s. of eq. (49). Owing to eqs
(41), (46) and (48) the result is identified with the curvature tensor
$R_{\underline \beta \alpha  \rho}^{\ \ \ \ \sigma}$  where $\alpha$ and
$\underline \beta$ are bosonic indices. Then we take the trace over $\rho$ and
$\sigma$. Remarkably the infinite sum  converges:
$$
\eqalign{
\sum_{{\rm bosonic}\ \sigma} R_{\underline \beta \alpha \sigma}^{\ \ \ \
\sigma} & = \sum_{{\rm fermionic}\ \sigma} R_{\underline \beta \alpha
\sigma}^{\ \ \ \ \sigma} \cr
& = {26\over 12}(\alpha^3 - {1\over 13}\alpha)\delta_{\alpha + \beta, 0} ,
\cr}
$$
as has been shown in ref. 8.
Note that it does not depend on the function $f(\alpha)$ at all. Consequently
we obtain the Ricci tensor
$$
R_{\alpha \underline \beta} = -R_{\underline \beta \alpha} =
-\sum_{{\rm all}\ \sigma} (-)^\sigma R_{\underline \beta \alpha \sigma}^{\ \ \
\ \sigma} = 0
$$
with bosonic indices $\underline \beta$ and $\alpha$.

\vskip 1 cm

\noindent
8.~~~~~~In this note we have given an explicit construction of the anti-bracket
of the BV formalism. The base space of the BV formalism is the fermionic
Virasoro manifold. The Ricci tensor, whose form curiously coincided with the
central charge of the Virasoro algebra in the bosonic $Diff(S^1)/S^1$ , turned
out to be vanishing in the fermionic one. We have studied the master equation
of the BV formalism and found an infinite number of Virasoro invariant
solutions, eq. (33).
The physical meaning of these solutions is that they could be physical states
of the topological $\sigma$-model$^{[11]}$ on the (bosonic) $Diff(S^1)/S^1$\
$^{[12]}$.

Finally we would like to remark that the anti-bracket having the Kac-Moody
symmetry can be similarly constructed along the arguments in this note.
Non-linear realization of the Kac-Moody algebra is necessary to do this. It has
been done in ref. 13.

\vskip 1 cm

\noindent
{\bf Acknowledgments}

The author thanks the Research Council of K.U. Leuven for the financial
support.

\vskip 1 cm

\noindent
{\bf References}

\noindent
\item{1.} I.A. Batalin and G.A. Vilkovisky, Phys. Lett. B102(1981)27; Phys.
Rev. D28(1983)2567.
\item{2.} E. Verlinde, Nucl Phys. B381(1992)141.
\item{3.} E. Witten, ``On background in\-de\-pen\-dent open-string field
the\-ory",
 IASSNS\--HEP\--92/\-53, hep-th/9208027, August 1992; ``Some computations in
background in\-de\-pen\-dent open-string field the\-ory", IASSNS\--92/\-93,
hep\--th/\-9210
 \item{}065, Oc\-to\-ber 1992;
\item{} H. Hata and B. Zwiebach, ``Developing the covariant Batalin-Vilkovisky
approach to the string theory", MIT-CTP-2177, hep-th/9301097, January 1993.
\item{4.} E. Witten, Mod. Phys. Lett. A5(1990)487;
\item{5.} A. Schwarz, ``Geometry of Batalin-Vilkovisky quantization ", UC Davis
preprint, hep-th/9205088, May 1992.
\item{6.} S. Aoyama and S. Vandoren, ``The Batalin-Vilkovisky for\-mal\-ism on
ferm\-ionic K\"ahler manifolds", KUL-TF-93/15, hep-th/9305087, May 1993.
\item{7.} O.M. Khudaverdian and A.P. Nersessian, ``On the geometry of the BV
formalism", UGVA-93/03-807,  March 1993.
\item{8.} J. Wess, ``Non-linear realization, K\"ahler manifolds and the
Virasoro manifold", lecture in 7th Scheveningen Conf. (August 1987);
\item{} B. Zumino, ``The geometry of the Virasoro group for physicists, lecture
in Carg\`ese Summer School (August 1987).
\item{9.} C.G. Callan, S. Coleman, J. Wess and B. Zumino, Phys. Rev.
177\-(1969)\-2247.
\item{10.} M.J. Bowick and S.G. Rajeev, Nucl. Phys. B293(1987)348.
\item{11.} E. Witten, Comm. Math. Phys. 118(1988)411.
\item{12.} S. Aoyama, ``Quantization of the topological $\sigma$-model and the
master equation of the BV formalism", KUL-TF-93/39, hep-th/9309103, September
1993.
\item{13.} S. Aoyama, Phys. Lett. B207(1988)130.

\bye